\documentclass[twocolumn,showpacs,preprintnumbers,amsmath,amssymb]{revtex4}

\usepackage{graphicx}
\usepackage{dcolumn}
\usepackage{bm}

\begin{document}

\preprint{APS/123-QED}

\title{Spin dynamics in Heisenberg triangular antiferromagnets: A $\mu$SR study of LiCrO$_2$}

\author{A. Olariu$^{1,*}$}
\author{P. Mendels$^1$}
\author{F. Bert$^1$}
\author{L. K. Alexander$^{1,2}$}
\author{A. V. Mahajan$^2$}
\author{A. D. Hillier$^3$}
\author{A. Amato$^4$}

\affiliation{$^1$Laboratoire de Physique des Solides, UMR 8502 CNRS, Universit\'e Paris-Sud, 91405 Orsay, France}

\affiliation{$^2$Department of Physics, Indian Institute of Technology Bombay, Mumbai 400076, India}

\affiliation{$^3$ISIS Facility, Rutherford Appleton Laboratory, Chilton, Didcot, Oxon OX11 OQX, United Kingdom}

\affiliation{$^4$Laboratory for Muon Spin Spectroscopy, Paul Scherrer Institute, CH-5232 Villigen PSI, Switzerland}

\begin{abstract}
We report a $\mu$SR study of LiCrO$_2$, which has a magnetic lattice made up of a stacking of triangular Heisenberg antiferromagnetic (Cr$^{3+}$, $S = 3/2$) layers. A static magnetically ordered state is observed below the transition temperature $T_N= 62$~K, while the expected peak of the relaxation rate is slightly shifted downward by a few kelvins below $T_N$. We draw a comparison with the isostructural compound NaCrO$_2$, where an exotic broad fluctuating regime has been observed [A. Olariu, P. Mendels, F. Bert, B. G. Ueland, P. Schiffer, R. F.
Berger, and R. J. Cava, Phys. Rev. Lett. \textbf{97}, 167203 (2006)] and was suggested to originate from topological excitations of the triangular lattice. Replacing Na by Li strongly narrows the exotic fluctuating regime formerly observed in NaCrO$_2$, which we attribute to a more pronounced inter-plane coupling in LiCrO$_2$.
\end{abstract}

\maketitle

\section{Introduction}

Geometrical frustration of antiferromagnetic interactions is the object of intense research as a route to generate novel exotic ground states. First proposed by Anderson for the triangular lattice~\cite{Anderson73}, spin-liquid-like behavior is now commonly observed in many compounds with corner-shared triangular lattices. Archetypes are the kagome or the pyrochlore lattices~\cite{Mendels07, Helton07,Gardner99}. The surge of interest for such weakly connected lattices in the 1990s came from (i) the lack of good experimental realizations of Heisenberg triangular antiferromagnets and (ii) the idea that a weaker connectivity offers an additional ingredient to destabilize N\'eel states and generate novel exotic physics.

Only very recently, the discovery of new experimental systems revived the interest in the pure triangular lattice. From the theoretical point of view, there was an early consensus that the Heisenberg triangular antiferromagnet with nearest-neighbor coupling orders at $T=0$ in the so-called 120$^{\circ}$ configuration of spins~\cite{TheseLecheminant}. However, to date, there has been little theoretical exploration in the dynamics associated with the triangular topology. Kawamura and Miyashita proposed in the 1980s that, even though no magnetic transition is possible at a finite temperature, the system goes through a cross-over regime due to the proliferation of topological defects Z$_2$~\cite{Kawamura84}. Observations of exotic fluctuating behaviors are at the heart of the recent experimental explorations of the triangular lattice since they might have some connection with the aforementioned topological excitations. In NiGa$_2$S$_4$, a fluctuating behavior is observed down to low temperatures as compared to the exchange interaction~\cite{Nakatsuji05,Kawamura07} and recently a transition to an ordered state was finally observed at $\sim10$~K~\cite{Takeya08}. The case of Cs$_2$CuCl$_4$~\cite{Coldea02} is more controversial. Initially thought to be an anisotropic triangular lattice, the fluctuating character might be due to a very pronounced one-dimensional (1D) character. In the two previous examples, the interaction paths are not straightforward. NaCrO$_2$, re-discovered in the context of the surge of interest for Na cobaltates,  offered a third, much simpler route to tackle the issue of the nature of the ground state. As shown in ~\cite{Olariu06}, it features well-decoupled perfect Heisenberg triangular planes. A broad fluctuating regime was observed well below the temperature at which a peak was observed in the magnetic susceptibility and was tentatively associated with topological excitations such as Z$_2$ vortices.

In this paper, we present results on another stable member of the ACrO$_2$ family (A = Li, Na, K), namely, LiCrO$_2$. Our main motivation is to probe the changes in the exotic dynamics observed in NaCrO$_2$ through the modification of the structural parameters and, hence, a modification of the magnetic couplings.

\section{AC$\textrm{r}$O$_2$ (A=L$\textrm{i}$, N$\textrm{a}$, K) family}

As reviewed by Collins and Petrenko, most of the studied triangular compounds had been far from ideal since they presented either a large anisotropy of the interactions or a sizeable single-ion anisotropy and/or next-neighbor coupling~\cite{Collins97}. To cite some of them, compounds from the family $ABX_3$, where $A$ is an alkali metal, $B$ is a transition metal, and $X$ is a halogen atom, were very well studied in the past. They have quasi-1D character, with coupling between planes much larger than the intra-plane one. Closer to the ideal Heisenberg case are the systems that belong to the family V$X_2$, where $X$=Cl, Br, and I. However, in these compounds single-ion anisotropy was observed while the triangular planes are weakly coupled.

\begin{figure}
\includegraphics[width=\linewidth]{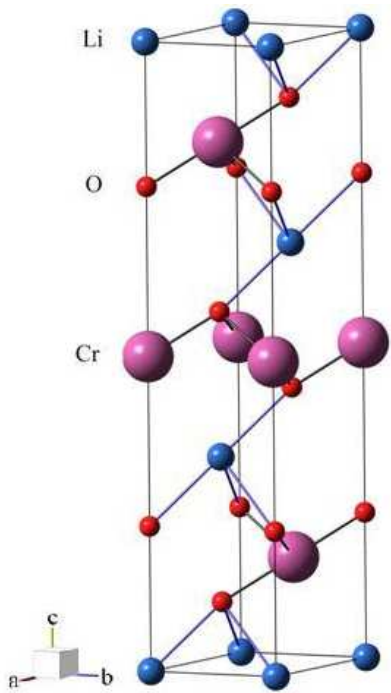}
\caption{The LiCrO$_2$ structure is built of triangular planes of Li, O and Cr in an $ABCABC$ sequence.}
\label{Structure}
\end{figure}

In the ACrO$_2$ family, the magnetic structure is built by stacking well-separated, triangular planes of Cr$^{3+}$ with spin $S=3/2$ in an $ABCABC$ sequence. A schematic structure is given in Fig.~\ref{Structure}. The R$\bar{3}$m symmetry of the structure ensures the isotropy of the in-plane-exchange coupling. Also, at least in the first two members of the series, this coupling is restricted to the nearest-neighbors~\cite{Mazin07,Olariu06}. Electron Spin Resonance (ESR) measurements indicate a very small single-ion anisotropy, thus revealing a very pronounced Heisenberg character which is quite common for Cr-oxides~\cite{Elliston75,Angelov84}. In this family, the $c$-lattice parameter increases sizably from 14.4~\AA  ~for Li to 17.9~\AA ~for K, which tunes the separation between Cr planes. Here the $c$ parameter is given for the rhombohedral unit cell.

An early macroscopic susceptibility study on a wide temperature range presented in Ref.~\cite{Delmas78} showed that all three compounds present antiferromagnetic correlations. For LiCrO$_2$, $\theta_{CW}=700$~K and $J=78$~K, while for NaCrO$_2$ $\theta_{CW}=290$~K and $J=40$~K.

Neutron-scattering measurements on a single crystal of LiCrO$_2$ ~\cite{Kadowaki95} show the presence of three-dimensional (3D) antiferromagnetic order that develops below $\sim 64$~K in a complicated 120$^{\circ}$ structure based on 18 sublattices. Accordingly, macroscopic susceptibility and specific-heat-capacity data show maxima around $T_N\sim62$~K~\cite{Alexander07}.
Recent first-principles calculations on LiCrO$_2$ indicated the existence of weak coupling between the planes, which is possibly antiferromagnetic in nature~\cite{Mazin07}. A $^7$Li NMR study indicated the presence, above 60 K, of a line shift, which demonstrates the existence of hyperfine coupling between Li and Cr, probably mediated by the oxygen~\cite{Alexander07}. Below this temperature, a wipe out of the signal occurs, probably because of the fast relaxation of nuclei in the critical regime around the phase transition.

In a recent NMR and $\mu$SR study on NaCrO$_2$, an unusual regime of excitations was revealed, likely related to the triangular nature of the magnetic structure~\cite{Olariu06}. Note that no shift was found by $^{23}$Na measurements on NaCrO$_2$ in the high-$T$ regime, which was attributed to the absence of hyperfine coupling between Na and Cr~\cite{Olariu06}. As we will detail further in this paper, the presence of shift on $^7$Li NMR spectra reveals a more substantial 3D coupling between layers in LiCrO$_2$. In this paper we present a similar $\mu$SR study of LiCrO$_2$. Due to the 10~ns-15$\mu$s window that allows detection of fast-relaxing sites, $\mu$SR enables a more complete study of LiCrO$_2$ over the whole temperature range, than in NMR. We compare the temperature dependence of the relaxation rate for both compounds and discuss the relation with the strength of the coupling between planes.

Our polycrystalline samples were synthesized by the stoichiometric mixture of Li$_2$CO$_3$ and Cr$_2$O$_3$, heated at 800$^{\circ}$C for 48 h, with one intermediate grinding~\cite{Alexander07}. X-ray measurements showed them to be single phased with lattice parameters $a=b=2.9$~\AA~ and $c=14.4$~\AA, in good agreement with previous results~\cite{Delmas78}.

\section{Experimental Results}

$\mu$SR is a sensitive technique for the study of magnetic properties at a microscopic level. It relies on the implantation of positive muons, that have a spin $S_{\mu}=1/2$ and a very large gyromagnetic ratio, $\gamma/2\pi=135.5$~MHz/Tesla. The muon spin is initially 100~\% polarized opposite to the direction of the beam. It is commonly assumed that in oxides $\mu ^+$ stop at $\sim 1$\AA~ from the O$^{2-}$ sites~\cite{Brewer91} and the muon spin feels the dipolar magnetic field created by the local environment of the sample. The measured quantity is the asymmetry of the muon as a function of time, that provides information about the static and dynamical magnetic behavior of the system under study.

Complementary measurements presented in this paper were performed  either at PSI (Switzerland) or at ISIS (United Kingdom) facilities. The continuous character of the beam at PSI enables one to get a high time resolution as needed at low $T$ whereas the pulsed beam of ISIS yields the possibility to follow in detail slow relaxing signals as needed for the dynamical study.

\begin{figure}
\includegraphics[width=\linewidth]{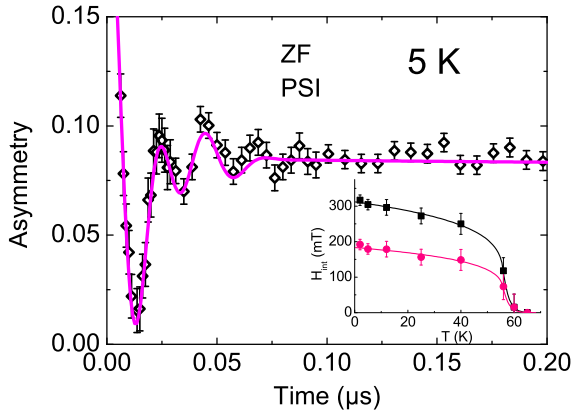}
\caption{Asymmetry in zero external field at early times and low temperature obtained for LiCrO$_2$. The spontaneous oscillation is a signature of an internal field indicating ordered freezing at this temperature.  The line is a fit to the data. Inset: Temperature dependance of the magnetic internal field, as extracted from the frequency of the oscillations. Lines are guides for the eye. }
\label{ZF_5Kzoom}
\end{figure}

The typical $\mu$SR asymmetry in zero external field (ZF), well below $T_N$, such as detected at 5~K with high statistics, is presented in Fig.~\ref{ZF_5Kzoom}. The spontaneous oscillations visible up to $\sim0.1~\mu$s indicate the presence of well-defined internal magnetic fields in the sample, showing that at this temperature the system is frozen and ordered. This finding perfectly agrees with neutron evidence~\cite{Kadowaki95} for a magnetic long-range order which can be described with 18 sublattices of spins at 120$^{\circ}$. In line with this quite involved structure, one can expect many magnetically inequivalent muon sites. For each site, the muon will experience a well-defined internal field and will contribute to the asymmetry with a given frequency. After inspection of the signal of Fig.~\ref{ZF_5Kzoom}, we were led to a minimal fitting model corresponding to a double-peaked histogram of the internal fields. This leads to a two-frequencies fit with a damping for each frequency corresponding to the expected but unresolved multiplicity of magnetically inequivalent $\mu^+$ sites. The asymmetry was therefore fitted according to the formula~\cite{footnote}:

\begin{equation}
a=a_0\sum_{i=1,\ 2}f_i \Big(\frac{2}{3} \textrm{cos} (2\pi\nu_i t+\phi)e^{-(\Delta t)^2/2 }+\frac{1}{3}\Big ) e^{-(t/T_1)^{\alpha}}
\end{equation}

Here, $f_i$ represent the weight of the two frequency components, found to be around 0.5 by the fit at 5~K and then fixed at this value at all temperatures. $a_0$ stands for the initial asymmetry at $t=0$ when muons are fully polarized (typically $a_0=0.245$ at PSI and $a_0=0.3$ at ISIS).
The fit obtained at 5~K is represented as a continuous line in Fig.~\ref{ZF_5Kzoom}. We obtained 43 and 26~MHz for the two frequencies of oscillations, which correspond to ~317 and 192~mT for the internal fields at the two muon sites. Their evolution with temperature is presented in the inset of Fig.~\ref{ZF_5Kzoom}. The field values decrease from a saturation value at low temperature down to zero around 60 K, where the oscillation pattern becomes difficult to follow. The fast damping of the oscillations, taken into account by the Gaussian exponential in the fitting function, indicates a distribution of sites of width $\sim100$~mT, in agreement with a large number of inequivalent magnetic sites.

\begin{figure}
\includegraphics[width=\linewidth]{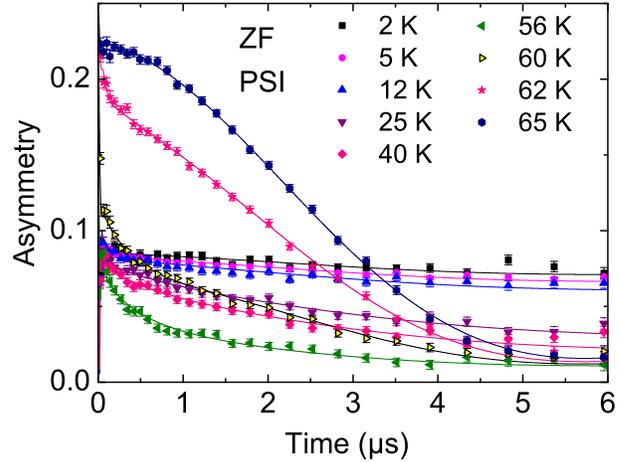}
\caption{Asymmetry obtained in zero external field for the whole studied temperature range. Lines are for fits, as described in the text.}
\label{ZF}
\end{figure}

We now turn to the study of dynamical effects, i.e., of electronic fluctuations, that can be tracked by the study of the asymmetry curves at long times. We used a combination of data from PSI and ISIS facilities. The advantage of the former is the absence of background due to muons not stopping into the sample, whereas the latter can be taken at long times with a good statistics which proved to be useful around the transition. Yet the uncertainty of the background values of the order of 0.005 is responsible for error bars in the relaxation rate.

Spectra obtained up to $6~\mu$s at PSI in zero external field are presented in Fig.~\ref{ZF}.
In the case of a complete freezing,  $T_1^{-1} \rightarrow 0$, and one obtains the so-called
"one-third" tail at long times ($a \rightarrow a_0/3$). When dynamical effects are present, fluctuations induce a relaxation of this tail.
In the fitting formula, a stretched exponential with $\alpha \sim0.5$ was used empirically to account for the distribution of magnetically inequivalent muon sites as mentioned above, which yield a distribution of relaxation rates.

At 5~K, the relaxation due to electronic fluctuations is small, $T_1^{-1}\sim0.03~\mu$s$^{-1}$, which shows that the behavior is mainly static.
As the temperature increases, the relaxation becomes more important. The one-third tail relaxation increases progressively up to $T\sim56$~K. At 56~K, the tail has the fastest decrease toward zero and, therefore, corresponds to the maximum of relaxation.

In the opposite limit of the fast fluctuating regime, above 65~K, the early time depolarization of the muon spin is mainly due to static nuclear dipoles, which create a small random field of $\sim$0.4~mT, independent of temperature. The induced asymmetry has the well-established Kubo-Toyabe (KT) shape, which is Gaussian-like on the time scale of Fig.~\ref{ZF}. In order to keep track of the slight relaxation of electronic origin, the KT function was multiplied by an exponential $exp(-\lambda t)$ with a weak value of $\lambda\sim0.05~\mu$s$^{-1}$.

\begin{figure}
\includegraphics[width=\linewidth]{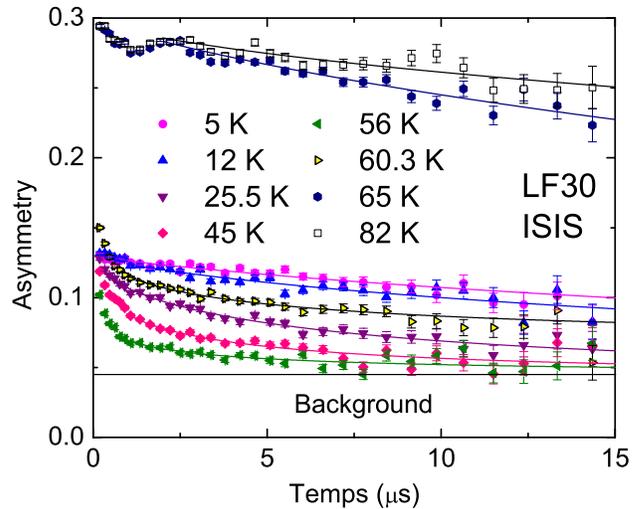}
\caption{Asymmetry obtained for an applied longitudinal field of 3~mT used to decouple the nuclear static contribution. Lines are for fits (see text).}
\label{LF}
\end{figure}

In the intermediate regime between 56 and 65~K, the data could not be fitted using the previous approach and we were led to conclude that two types of clusters, either static or dynamical, coexist. In order to simplify the fitting procedure, we performed experiments under a weak magnetic field of 3~mT, applied along the initial direction of the muon spin. Thus, the muon is no longer depolarized by the static field of the nuclear dipoles, which is much smaller than the applied field. On the other hand, this value is small enough so that it does not impact the electronic fluctuations. Asymmetry obtained at the ISIS facility is represented in Fig.~\ref{LF}. We fitted the asymmetry for $t > 1~\mu$s according to the ansatz

\begin{displaymath}
a=
\left\{ \begin{array}{l}
(a_0-B)\frac{1}{3}e^{-(t/T_1)^{\alpha}}+B\textrm{,\ for }T\leq56~\textrm{K}\\
\\
(a_0-B-a_p)\frac{1}{3}e^{-(t/T_1)^{\alpha}}+B+a_p,\\
\qquad\qquad\qquad\quad\textrm{for }56<T<65~\textrm{K}\\
\\
(a_0-B)e^{-(t/T_1)^{\alpha}}+B\textrm{,\ for }T\geq65~\textrm{K}\\

\end{array}
\right.
\end{displaymath}

Here the term $B$ stands for the background due to muons that do not stop into the sample ($B=0.045 \pm 0.005$). The term $a_p$ was introduced empirically in order to track the paramagnetic islands that survive below $T_N$, which adds to the background $B$. For simplicity, the relaxation corresponding to these sites was crudely neglected. The relaxation rate estimated for different values of $a_p$ varying from 0 to 0.035 were used in order to calculate the error bars. The obtained fits are represented as continuous lines in Fig~\ref{LF}. The temperature dependence of the relaxation rate is presented in Fig.~\ref{RelaxationRate}, and values obtained below 56~K are in good agreement with those obtained using the zero field spectra.

\section{Discussion}

\begin{figure}[t]
\includegraphics*[width=\linewidth]{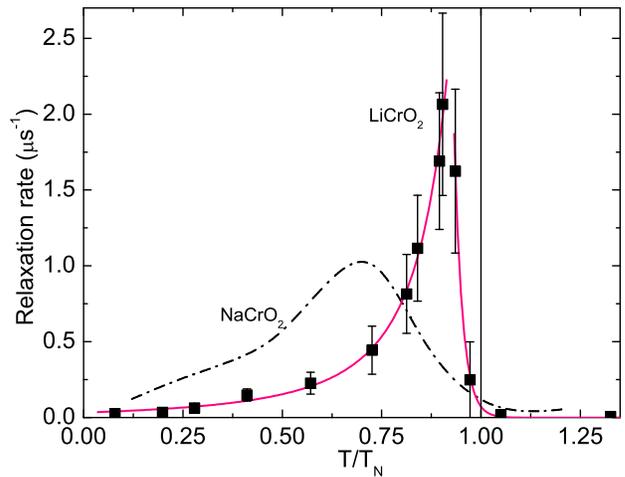}
\caption{Relaxation rate as a function of $T/T_N$ for NaCrO$_2$ and LiCrO$_2$. Lines are guides for the eye.}
\label{RelaxationRate}
\end{figure}

The temperature dependence of the relaxation rate is represented in Fig.~\ref{RelaxationRate} and compared to the Na analog. The temperature scale is normalized by $T_N$ (62~K for LiCrO$_2$, and 41~K for NaCrO$_2$). According to this figure, one can distinguish two features:

(i) in LiCrO$_2, $the relaxation rate peak appears at $\sim0.9 T_N$, while in NaCrO$_2$ it appears at $\sim0.75T_N$.

(ii) in LiCrO$_2$, the relaxation rate peak is sharper than for NaCrO$_2$.

Note that in compounds that present a regular transition, the relaxation-rate peaks where the specific heat does, and the width of the relaxation-rate peak is much smaller than in the case of NaCrO$_2$, typically 20\% of $T_N$~\cite{Coldea01_muSR,Maegawa07}.

In NaCrO$_2$, the relaxation rate cross-over is very broad and appears well below $T_N$, which indicates the presence of an \emph{extended fluctuating regime}. Accordingly, neutron-scattering results~\cite{Hsieh08} show no sign of 3D ordering above $\sim0.75T_N$ but only the presence of 2D correlations. Below this temperature, an incommensurate 3D magnetic order develops but remains short ranged down to lowest temperature.

In LiCrO$_2$, this unusual fluctuating regime appears to be very reduced in comparison with the Na case. The neutron-scattering study revealed the growing of 3D long-range magnetic order very close to the transition temperature~\cite{Kadowaki95}. Replacing Na by Li reduces the separation between successive Cr planes, which is likely to create a more important coupling in LiCrO$_2$. Note that in both Na and Li compounds the dominant mechanism for in-plane coupling is made by \emph{direct} exchange, as shown by band-structure calculations~\cite{Mazin07}, as well as by calculations that take into account the value of the Cr-O-Cr angle~\cite{Motida70}. This angle varies only little, from 96.4$^{\circ}$ in NaCrO$_2$ to 98.8$^{\circ}$ in LiCrO$_2$. In fact only minor changes occur in the CrO$_2$ planes which cannot yield the observed differences between the two compounds.

The scenario of a 3D coupling in LiCrO$_2$ is further supported by $^7$Li NMR measurements~\cite{Alexander07}, which revealed the presence of a line shift, implying the existence of hyperfine coupling between Li and Cr, probably mediated by the oxygen. The coupling between adjacent Cr$^{3+}$ planes can therefore proceed indirectly by the paths Cr-O-Li-O-Cr or alternatively Cr-O-O-Cr. The coupling through direct overlapping of Cr$^{3+}$ orbitals is probably negligible, in view of the large separation between the planes, of about 4.8~\AA. In addition to the exchange interaction, the Cr$^{3+}$ planes are also coupled by a small dipolar interaction of less than 1~K. On the contrary, in the Na compound, the $^{23}$Na NMR study revealed the absence of line shift, and by comparison with results on the isostructural Na$_x$CoO$_2$, it was inferred that there is no overlapping between the Na and O orbitals~\cite{Olariu06}. The main mechanisms for coupling, via the paths Cr-O-Na-O-Cr or Cr-O-O-Cr, are therefore not effective. The inter-layer coupling appears thus to be mainly dipolar in nature, so extremely small.

The comparative study of Li and Na Chromates allows, thus, to track the influence of the 3D coupling on the magnetic properties and isolate those intrinsic to the triangular lattice. In NaCrO$_2$, it was suggested that the extended fluctuating regime is a signature of topological defects, Z$_2$ vortices, whose existence was proposed long ago by Kawamura and Miyashita~\cite{Kawamura84}. The dissociation of vortex pairs would induce a transition at a temperature lower than that of the specific-heat-capacity peak. The present comparison underlines the sensitivity of such a mechanism to 3D couplings, and the sensitivity to similar deviations from the ideal 2D Heisenberg case might explain why the experimental signature of these defects has not been observed in real systems for a long time. Altogether with the recent discovery of NiGa$_2$S$_4$, this opens a new route to the exploration of the physics of edge-sharing triangular Heisenberg lattices. Indeed, in NiGa$_2$S$_4$, where Ni$^{2+}$ ions with $S=1$ form a triangular lattice, persisting dynamics was observed down to very low temperature, well below the specific-heat peak~\cite{Nakatsuji05,Takeya08} and the magnetic order revealed by neutron scattering measurements remains short ranged down to 0.35~K. Along these  ideas, it would be very interesting to
study KCrO$_2$, if this compound can be stabilized, for which one expects the most pronounced 2D character in the ACrO$_2$ family, since it has the largest separation between Cr planes. An early neutron-scattering work interestingly revealed the absence of any 3D magnetic ordering down to 2~K~\cite{Soubeyroux79}.\\

\section{Conclusion}

Our $\mu$SR study shows that the triangular Heisenberg compound LiCrO$_2$ develops a more conventional ordering than NaCrO$_2$, although the fluctuating regime remains sizeable. We propose an interpretation in terms of a more pronounced 3D coupling between CrO$_2$ triangular planes related to the smaller distance between them. This point is in line with neutron experiments and recent NMR results which overall reveals the importance of the 2D character in the study  of topological excitations typical of triangular systems.

\begin{acknowledgments}
This work is supported by ANR under Grant No. NT05-4-41913 "OxyFonda". We gratefully acknowledge support from the European Commission under Framework Programme 6 through the Key
Action: Strengthening the European Research Area, Research
Infrastructures under Contract No. RII3-CT-2003-505925.
\end{acknowledgments}


\begin{thebibliography}{24}
\bibitem{footnote}
\bibinfo{author}{To fit the PSI data, we added to expression (1) a temperature independent fast relaxing signal corresponding to a field distribution width $\simeq$ 600~G in order to take into account a $\simeq$ 10\% macroscopically segregated parasitic phase, which is magnetically frozen well above 65~K and which gives a fast decay at early times.}
\end{thebibliography}

\end{document}